\documentclass[aps,prd,twocolumn,superscriptaddress,showpacs,floatfix,nofootinbib]{revtex4}

\usepackage[dvips]{graphicx}
\usepackage{xspace}
\usepackage{latexsym}
\usepackage{amssymb}
\usepackage{times}
\usepackage{mathptm}
\usepackage{subfigure}

\begin{document}

\title{Search for Proton Decay via $p\rightarrow\mu^+K^0$ in Super-Kamiokande~I, II, and III}

\newcommand{\AFFicrr}{\affiliation{Kamioka Observatory, Institute for Cosmic Ray Research, University of Tokyo, Kamioka, Gifu 506-1205, Japan}}
\newcommand{\AFFkashiwa}{\affiliation{Research Center for Cosmic Neutrinos, Institute for Cosmic Ray Research, University of Tokyo, Kashiwa, Chiba 277-8582, Japan}}
\newcommand{\AFFipmu}{\affiliation{Kavli Institute for the Physics and
Mathematics of the Universe (WPI), Todai Institutes for Advanced Study,
University of Tokyo, Kashiwa, Chiba 277-8583, Japan }}
\newcommand{\AFFmad}{\affiliation{Department of Theoretical Physics, University Autonoma Madrid, 28049 Madrid, Spain}}
\newcommand{\AFFbu}{\affiliation{Department of Physics, Boston University, Boston, MA 02215, USA}}
\newcommand{\AFFbnl}{\affiliation{Physics Department, Brookhaven National Laboratory, Upton, NY 11973, USA}}
\newcommand{\AFFuci}{\affiliation{Department of Physics and Astronomy, University of California, Irvine, Irvine, CA 92697-4575, USA }}
\newcommand{\AFFcsu}{\affiliation{Department of Physics, California State University, Dominguez Hills, Carson, CA 90747, USA}}
\newcommand{\AFFcnm}{\affiliation{Department of Physics, Chonnam National University, Kwangju 500-757, Korea}}
\newcommand{\AFFduke}{\affiliation{Department of Physics, Duke University, Durham NC 27708, USA}}
\newcommand{\AFFfukuoka}{\affiliation{Junior College, Fukuoka Institute of Technology, Fukuoka, Fukuoka 811-0295, Japan}}
\newcommand{\AFFgmu}{\affiliation{Department of Physics, George Mason University, Fairfax, VA 22030, USA }}
\newcommand{\AFFgifu}{\affiliation{Department of Physics, Gifu University, Gifu, Gifu 501-1193, Japan}}
\newcommand{\AFFuh}{\affiliation{Department of Physics and Astronomy, University of Hawaii, Honolulu, HI 96822, USA}}
\newcommand{\AFFkanagawa}{\affiliation{Physics Division, Department of Engineering, Kanagawa University, Kanagawa, Yokohama 221-8686, Japan}}
\newcommand{\AFFkek}{\affiliation{High Energy Accelerator Research Organization (KEK), Tsukuba, Ibaraki 305-0801, Japan }}
\newcommand{\AFFkoba}{\affiliation{Kobayashi-Maskawa Institute for the Origin of Particles and the Universe, Nagoya University, Nagoya, Aichi 464-8602, Japan}}
\newcommand{\AFFkobe}{\affiliation{Department of Physics, Kobe University, Kobe, Hyogo 657-8501, Japan}}
\newcommand{\AFFkyoto}{\affiliation{Department of Physics, Kyoto University, Kyoto, Kyoto 606-8502, Japan}}
\newcommand{\AFFumd}{\affiliation{Department of Physics, University of Maryland, College Park, MD 20742, USA }}
\newcommand{\AFFmit}{\affiliation{Department of Physics, Massachusetts Institute of Technology, Cambridge, MA 02139, USA}}
\newcommand{\AFFmiyagi}{\affiliation{Department of Physics, Miyagi University of Education, Sendai, Miyagi 980-0845, Japan}}
\newcommand{\AFFnagoya}{\affiliation{Solar Terrestrial Environment Laboratory, Nagoya University, Nagoya, Aichi 464-8602, Japan}}
\newcommand{\AFFpol}{\affiliation{National Centre For Nuclear Research, 00-681 Warsaw, Poland}}
\newcommand{\AFFsuny}{\affiliation{Department of Physics and Astronomy, State University of New York, Stony Brook, NY 11794-3800, USA}}
\newcommand{\AFFniigata}{\affiliation{Department of Physics, Niigata University, Niigata, Niigata 950-2181, Japan }}
\newcommand{\AFFokayama}{\affiliation{Department of Physics, Okayama University, Okayama, Okayama 700-8530, Japan }}
\newcommand{\AFFosaka}{\affiliation{Department of Physics, Osaka University, Toyonaka, Osaka 560-0043, Japan}}
\newcommand{\AFFseoul}{\affiliation{Department of Physics, Seoul National University, Seoul 151-742, Korea}}
\newcommand{\AFFshizuokasc}{\affiliation{Department of Informatics in
Social Welfare, Shizuoka University of Welfare, Yaizu, Shizuoka, 425-8611, Japan}}
\newcommand{\AFFskk}{\affiliation{Department of Physics, Sungkyunkwan University, Suwon 440-746, Korea}}
\newcommand{\AFFtohoku}{\affiliation{Research Center for Neutrino Science, Tohoku University, Sendai, Miyagi 980-8578, Japan}}
\newcommand{\AFFtokyo}{\affiliation{The University of Tokyo, Bunkyo, Tokyo 113-0033, Japan }}
\newcommand{\AFFtokai}{\affiliation{Department of Physics, Tokai University, Hiratsuka, Kanagawa 259-1292, Japan}}
\newcommand{\AFFtit}{\affiliation{Department of Physics, Tokyo Institute
for Technology, Meguro, Tokyo 152-8551, Japan }}
\newcommand{\AFFtsinghua}{\affiliation{Department of Engineering Physics, Tsinghua University, Beijing, 100084, China}}
\newcommand{\AFFuw}{\affiliation{Department of Physics, University of Washington, Seattle, WA 98195-1560, USA}}

\AFFicrr
\AFFkashiwa
\AFFmad
\AFFbu
\AFFbnl
\AFFuci
\AFFcsu
\AFFcnm
\AFFduke
\AFFfukuoka
\AFFgifu
\AFFuh
\AFFkek
\AFFkobe
\AFFkyoto
\AFFmiyagi
\AFFnagoya
\AFFkoba
\AFFpol
\AFFsuny
\AFFokayama
\AFFosaka
\AFFseoul
\AFFshizuokasc
\AFFskk
\AFFtokai
\AFFtokyo
\AFFipmu
\AFFtsinghua
\AFFuw

\author{C.~Regis} 
\AFFuci

\author{K.~Abe}
\AFFicrr
\author{Y.~Hayato}
\AFFicrr
\AFFipmu
\author{K.~Iyogi} 
\author{J.~Kameda}
\author{Y.~Koshio}
\author{Ll.~Marti}
\author{M.~Miura} 
\AFFicrr
\author{S.~Moriyama} 
\author{M.~Nakahata} 
\AFFicrr
\AFFipmu
% ...

\collaboration{The Super-Kamiokande Collaboration}
\noaffiliation

\date{\today}

\begin{abstract}

We have searched for proton decay via $p\rightarrow\mu^+K^0$
using data from a 91.7~kiloton$\cdot$year exposure of Super-Kamiokande-I,
a 49.2~kiloton$\cdot$year exposure of Super-Kamiokande-II,
and a 31.9~kiloton$\cdot$year exposure of Super-Kamiokande-III.
The number of candidate events in the data was consistent with
the atmospheric neutrino background expectation and no evidence for 
proton decay in this mode was found.
We set a partial lifetime lower limit of 1.6$\times10^{33}$~years
at the 90\% confidence level.

\end{abstract}

\maketitle

\section{INTRODUCTION}

The experimental observation of proton decay
would provide strong evidence for Grand Unified Theories (GUTs)
~\cite{guts} which unify the strong, electromagnetic, and weak forces. 
Supersymmetry (SUSY) GUTs predict the unification of coupling constants
consistent with the non-observation of $p\rightarrow e^+\pi^0$~\cite{skepi0}
via gauge boson exchange.
In some SUSY GUT models~\cite{nuk}
$p\rightarrow{\bar\nu}K^+$ is then favored by dimension five operators.
However, in other models~\cite{theories}
$p\rightarrow\mu^+K^0$ may dominate.
So far there has been no experimental observation of either of these decays
and in this paper we focus on the later, whose most stringent partial 
lifetime lower limit comes from Super-Kamiokande~(SK), 
$\tau$/$B_{p\rightarrow\mu^+K^0}$ $>$ 1.3$\times$10$^{33}$~years
(90\% C.L.)~\cite{kk}.
The partial lifetime of $p\rightarrow\mu^+K^0$  
is predicted to be just above this limit \cite{theories}.

The $K^0$ is a composite state of the $K^0_S$ (50\%) and $K^0_L$ (50\%).
The analysis presented in \cite{kk} searched only for proton decay 
into $\mu^+K^0_S$ using data from the first run period of SK
(91.7 kton$\cdot$years).
In this paper we present a combined search for proton decay into 
both $K^{0}_S$ and $K^{0}_L$ with an updated data set 
(172.8 kton$\cdot$years).
 
A $p\rightarrow\mu^+ K^0$ signal would appear in SK
as one muon-like ($\mu$-like) ring 
with monochromatic momentum of about 327~MeV/$c$
and secondary ring(s) from the decay of the $K^0$.
The total momenta of all such rings should be
consistent with the decay of a proton, as well as an invariant mass
close to the proton mass.
The dominant decay modes of the $K^0_S$ are $\pi^+\pi^-$ (69.2\%),
and $\pi^0\pi^0$ (30.7\%).
The $K^0_L$ decays predominantly into 
$\pi^{\pm}e^{\mp}\nu_{e}$ (40.6\%),
$\pi^{\pm}\mu^{\mp}\nu_{\mu}$ (27.0\%),
$3\pi^{0}$ (19.5\%),
and $\pi^{+}\pi^{-}\pi^{0}$ (12.5\%).
Since the $K^0_L$ has a relatively long decay length
($c\beta\tau\sim$8.4~m, where $\tau$ is the mean life),
proton decay events into this channel are characterized 
by two vertices separated in time and space: a primary vertex from which 
the $\mu^{+}$ emerges and a secondary vertex for the decay products of the $K^{0}_L$.
We developed new event reconstruction for events with two vertices.

The outline of this paper is as follows.
A review of the SK detector is presented in Section~II.
In section~III the $p\rightarrow\mu^+ K^0$ Monte Carlo (MC)
simulation as well as the atmospheric neutrino MC simulation
used to calculate the signal efficiency and expected background rate
are presented.
Section~IV and V describe the data reduction and the event reconstruction, 
respectively.
The results of the data analysis are summarized in Section~VI  
and a comparison with the previous result~\cite{kk} 
is described in subsection~VI-C.
Finally, Section~VII concludes this study.

\section{SUPER-KAMIOKANDE DETECTOR}

The Super-Kamiokande detector~\cite{skdet}
is located in Gifu Prefecture, Japan, at a 
depth of 2,700~m water equivalent within the 
Kamioka mine at Mt.~Ikenoyama. 
The detector is a 50~kton right cylinder
that is optically separated into two concentric regions, an inner (ID) and an outer detector (OD).
Inward-facing 20~inch diameter Hamamatsu photomultiplier tubes (PMTs)~\cite{pmt}
are mounted uniformly on the walls of the ID.
During the SK-I (SK-III) run period from April 1996 (October 2006) to July 2001 (September 2008)
approximately 11,100~PMTs lined the ID walls.
There were about 5,200~PMTs in the ID during the SK-II run period  
from October 2002 to October 2005.
The photocathode coverage during the run was about 40\% (20\%) 
in SK-I and -III (SK-II).
A black sheet spans the region between PMTs
to absorb light and optically separate ID and OD.
Starting with SK-II, each ID PMT is encased in an acrylic cover
to reduce the effect of implosion on neighboring PMTs. 
The transparency of this cover is more than 96\% 
for $>$350~nm photons with normal incidence. 
A 22.5~kton fiducial volume used in the data analysis presented below 
is defined by the cylindrical volume taken 2~m from the ID walls.  

The OD is a 2~m thick cylindrical shell enclosing the 
ID that is instrumented with 1,885 outward-facing 8~inch diameter Hamamatsu PMTs.
To improve the light collection efficiency,
a 60 cm $\times$ 60 cm wavelength shifting plate is attached
to each PMT and the OD walls are covered
with reflective material called ``Tyvek sheet''.
The main purpose of the OD is to tag incoming cosmic ray muons 
and muons induced by atmospheric neutrinos that escape the ID.
The OD region also serves as a passive shield
against radioactivity from outside the detector wall.
The SK-I detector and its calibration are described 
in detail elsewhere \cite{skdet}.

\section{SIMULATION}

The $p\rightarrow\mu^+ K^0$ MC
and the SK atmospheric neutrino MC
are used to estimate signal efficiencies
and the expected background rates.
Using these MCs 
the event selections described in Sec.~VI-A are optimized 
to maximize the analysis' sensitivity prior to examining the data.

\subsection{Proton Decay}

The $p\rightarrow\mu^+ K^0$ MC used here is the same as that used in \cite{kk}
with the exception that $p\rightarrow\mu^+ K^0_L$ events
are additionally simulated in this analysis.
In this analysis the decays of both free proton
and protons bound on the oxygen nucleus are considered. 
For the decay of free protons the directions of the 
$\mu^+$ and $K^0$ are exactly back-to-back, and each
has a momentum of 326.5~MeV/$c$. 
In the case of proton decay within the oxygen nucleus, 
the effects of Fermi momentum, correlation with other nucleons, 
the nuclear binding energy, kaon-nucleon interactions, 
and $K^0_L\rightarrow K^0_S$ regeneration are all taken into account.
The Fermi momentum distribution 
measured by electron-$^{12}$C scattering~\cite{fermi}
is used in the simulation. 
The nuclear binding energy effect is taken into account 
by modification of the proton's mass.
The modified proton mass, $m_p'$, is calculated by
$m_p' = m_p - E_b$ where $m_p$ is the proton rest mass
and $E_b$ is the nuclear binding energy.
The value of $E_b$ for each simulated event was randomly
selected from a Gaussian probability density function
with ($\mu$, $\sigma$) = (39.0, 10.2) MeV for the S-state and
($\mu$, $\sigma$) = (15.5, 3.82) MeV for the P-state.
The correlation of the decaying proton's wave function 
with other nucleons reduces the invariant mass of the 
decay products since some of the system's momentum is carried 
by these nucleons. 
This effect is present in about 10\% of proton decays~\cite{corrdcy}
and produces a broad invariant mass spectrum down to about 600~MeV/$c^2$.
The location of protons within the $^{16}$O nucleus 
is calculated according to the Woods-Saxon nuclear density 
model~\cite{woodsax}.
The kaon nucleon interactions within the $^{16}$O nucleus 
that are considered include
elastic scattering and inelastic scattering via charge exchange.
The $K^+N$ scattering amplitudes were fitted
by a partial wave analysis using many data samples 
by Hyslop {\it et al.}~\cite{hyslop}.
The $K^+ n\rightarrow K^0 p$ charge exchange cross section 
was measured by Glasser {\it et al.}~\cite{glasser}.
The detail of the $K^+N$ interaction is described in \cite{kk}.
From isospin symmetry,
the $K^0N$ reactions have essentially the same magnitude
as the $K^+N$ reactions.
If an interaction occurs, the effect of Pauli blocking is introduced 
by requiring the nucleon momentum after interaction be larger than 
the Fermi surface momentum,
given by $p_F(r) = (\frac{3}{2}\pi^2\rho(r))^\frac{1}{3}$,
where $\rho(r)$ is the nuclear density distribution
and $r$ is the distance from the center of the nucleus.
The $K^0_L\rightarrow K^0_S$ regeneration effect 
is implemented as a decay mode of the $K^0_L$ in this analysis.
The regeneration probability in water
is based on the results of a kaon scattering experiment 
using a carbon target~\cite{regen}.
Regeneration within the $^{16}$O nucleus is also considered
and is assumed 
to be proportional to its density relative to water.
The fraction with regeneration is about 0.1$\%$
of the total generated $p\rightarrow\mu^+ K^0_L$ MC events.

We simulate propagation of particles 
and Cherenkov light production in water by custom software based on
GEANT~\cite{geant}. The propagation of charged pions in water
is simulated using custom code based on \cite{skpimc} for pion momenta less
than 500~MeV/$c$ and using CALOR~\cite{calor} otherwise.

\subsection{Atmospheric Neutrinos}

The interactions of atmospheric neutrinos within SK pose the 
dominant source of background to searches for nucleon decay.
For the $p \rightarrow \mu^{+} K^{0}$ mode, the decay products of 
the $K^{0}_{L}$ and $K^{0}_{S}$ often include pions which 
can be created in atmospheric neutrino interactions.
Notably, charged current multiple pion production 
and deep inelastic scattering processes, again with multiple 
pions in the final state, are backgrounds to the search presented here. 
Accordingly, the accurate simulation of atmospheric neutrino events 
in SK is essential to properly estimate the proton decay background.
In this paper, the atmospheric neutrino MC is the same as used in other 
SK analyses~\cite{skepi0,skatmosc}. The primary flux is taken 
from the Honda model~\cite{honda} and atmospheric neutrino interactions 
are generated within the detector 
using the NEUT simulator~\cite{neut,ktbgpaper}.
Kaon regeneration is not considered in the atmospheric neutrino MC 
because the fraction of $K^0$ production in these events is negligibly small.
For each of the SK run periods presented here, 
independent about 200 year MC samples 
were used to compute the background 
rates for the present search. Two-flavor neutrino oscillation effects with 
$\mbox{sin}^{2} 2\theta_{23} = 1.0$ and 
$\Delta m^{2} = 2.5 \times 10^{-3} \mbox{eV}^{2}$ 
are included in the background prediction. 
A more detailed explanation of the atmospheric neutrino MC 
is presented in \cite{skepi0,skatmosc}.

\section{DATA SET AND REDUCTION}

The current data set corresponds to a 91.7 kton$\cdot$years 
(1489.2~days) exposure 
from SK-I, a 49.2 kton$\cdot$years (798.6~days) exposure during SK-II,
and a 31.9 kton$\cdot$years (518.1~days) exposure from SK-III.
The data acquisition trigger threshold
used in this data analysis
during these periods 
corresponds to the energy deposited 
by 6~MeV electron during SK-I and SK-III
and 8~MeV electron during SK-II.
Most events above this threshold are cosmic ray muons or
low energy backgrounds from radioactivity in detector materials.
To remove these backgrounds the data are passed through several reduction stages 
before detailed reconstruction processes (described in Sec.~V) are applied.
To summarize, the starting point of the data analysis is the fully contained (FC) fiducial volume (FV) data sample,
which contains events passing the following cuts:
\begin{itemize}
\item the number of hit PMTs in the highest charge cluster of PMTs in the OD is less than 10
\item the total visible energy in the ID is greater than 30~MeV
\item the reconstructed vertex's distance to the ID wall is greater than 2~m 
\end{itemize}
After these cuts the FCFV event rate is about eight events per day.
Non-atmospheric neutrino backgrounds are estimated to represent less than 
1\% of this sample.  Details of the data reduction can be found in \cite{skatmosc}.

\section{EVENT RECONSTRUCTION}

Event reconstruction programs are applied
to the FCFV sample described in the previous section.
For the $K^0_S$ searches,
the standard event reconstruction~\cite{skatmosc,nimrec}
used in the SK atmospheric neutrino oscillation analyses
and proton decay searches is used.
However, relative to the previous version of this analysis~\cite{kk}, 
several improvements to the reconstruction algorithms have been made. 
The search for $K^{0}_L$ events makes 
use of new event reconstruction methods,
which are explained in detail below.

The event reconstruction for the $K^0_S$ searches is:

\begin{itemize}

\item vertex reconstruction

The event vertex is initially determined
by using the time of flight (TOF)-subtracted timing distribution
of hit ID PMTs assuming all light originated from a single point (point fit). 
An initial direction for the event is then estimated as the 
sum of vectors drawn from this vertex to each 
hit PMT and weighted by its observed charge. 
Defining this direction as the particle's trajectory, 
the distribution of observed charge as a function of 
the opening angle to this axis is used to determine
the edge of the brightest Cherenkov ring in the event. 
With the ring edge in hand, the vertex position and particle 
direction are refined using the timing residual distribution,
but assuming that PMTs within the ring are illuminated by 
photons generated along the particle track and other PMTs 
receive light generated at the event vertex.

\item ring counting

Based on the reconstructed vertex, 
additional ring candidates in the event are searched for
using a technique based on the Hough transform~\cite{hough}.
Possible ring candidates returned by the Hough mapping are tested  
using a likelihood method to determine if the existence of a second ring
better describes the observed PMT hit pattern than a single ring 
alone. If a second ring is found, the procedure is iterated 
up to a total of five rings. 
At each step the likelihood is based on a comparison of the observed 
charge in the event against the expectation from the number of 
proposed rings. With respect to the previous analysis there have been 
improvements to the probability density function (PDF) for the expected 
charge that are used in the present analysis. More details 
are presented below.

\item PID (e/$\mu$ separation)

After the number of rings has been determined each ring's 
particle type (PID) is estimated 
taking into account the effects of light scattering in the water
and reflection off PMT surfaces. 
Two likelihood functions are used to separate 
electromagnetic shower type ($e$-like) rings 
from muon type ($\mu$-like) rings. 
The first is a pattern matching algorithm that 
exploits topological differences in the PMT hit pattern 
that arise from the scattering and pair production processes
an electron experiences, but that the primarily unperturbed 
muon does not.
A second method makes use of the difference in the 
observed Cherenkov angle between the two particles 
as a result of their largely different masses.
For the final estimation of a ring's PID, the 
two methods are combined. 

\item decay electron search

Electrons produced from the decay of muons in the 
detector can occur in the same trigger timing gate with the primary event 
or delayed. The search for decay electrons extends 
to both cases.

\item momentum reconstruction

The momentum of each ring is 
determined using the sum of its assigned 
charges after correcting for  
the attenuation of Cherenkov light in water
and the incident angle dependence of the PMT acceptance.
Depending upon the result of the PID estimation, 
the momentum is then assigned based on the 
amount of corrected charge expected for the fitted particle 
type. A pionic momentum ($\pi$-like) that relies 
on the expected Cherenkov angle as well as the total 
corrected charge is also used in this analysis.
The time variation of the light attenuation length
and the PMT gain, as measured with cosmic rays
traversing the detector, 
is incorporated into the momentum reconstruction.

\end{itemize}

Improvements made to the single photoelectron response functions used 
to predict the expected charge in each PMT
have led to improvements in the ring counting, PID, and momentum estimations
for this analysis.

\begin{figure}[htbp]
\begin{center}
  \includegraphics[width=9cm,clip]{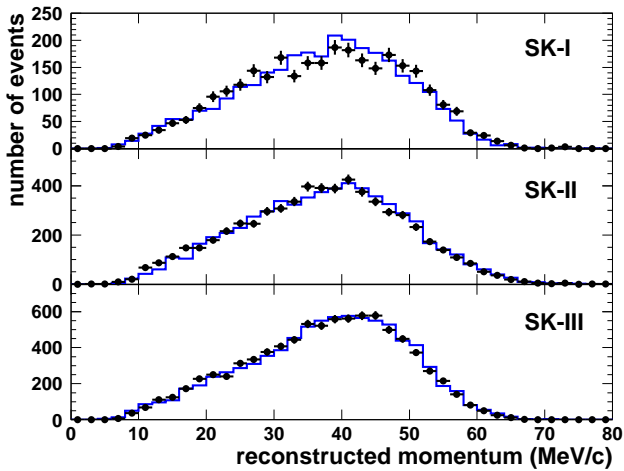}
  \caption
  {\protect \small 
The momentum distribution of decay electrons
of the data (dot) and MC (solid line)
for SK-I, SK-II, and SK-III from top to bottom, respectively.
MC events are normalized by the number of observed data events.
The mean values of the data distributions agree with those of the MC
within 0.6\%, 1.6\%, and 0.8\%
for SK-I, SK-II, and SK-III, respectively.}
  \label{decay-e}
\end{center}
\end{figure}

\begin{figure}[htbp]
\begin{center}
  \includegraphics[width=9cm,clip]{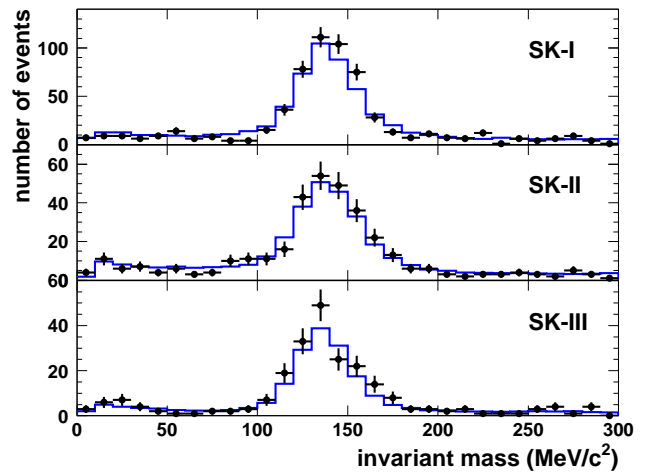}
  \caption
  {\protect \small 
Invariant mass distribution of neutrino-induced $\pi^0$ events
of the observed data (dot) and 
the atmospheric neutrino MC events (solid line)
for SK-I, SK-II, and SK-III from top to bottom, respectively.
MC events are normalized by the livetime of the observed data.
The peak positions of the data distributions agree with those from the MC
within 0.7\%, 1.3\%, and 0.3\%
for SK-I, SK-II, and SK-III, respectively.}
\label{pi0mass}
\end{center}
\end{figure}

\begin{figure}[htbp]
\begin{center}
  \includegraphics[width=\linewidth,clip]{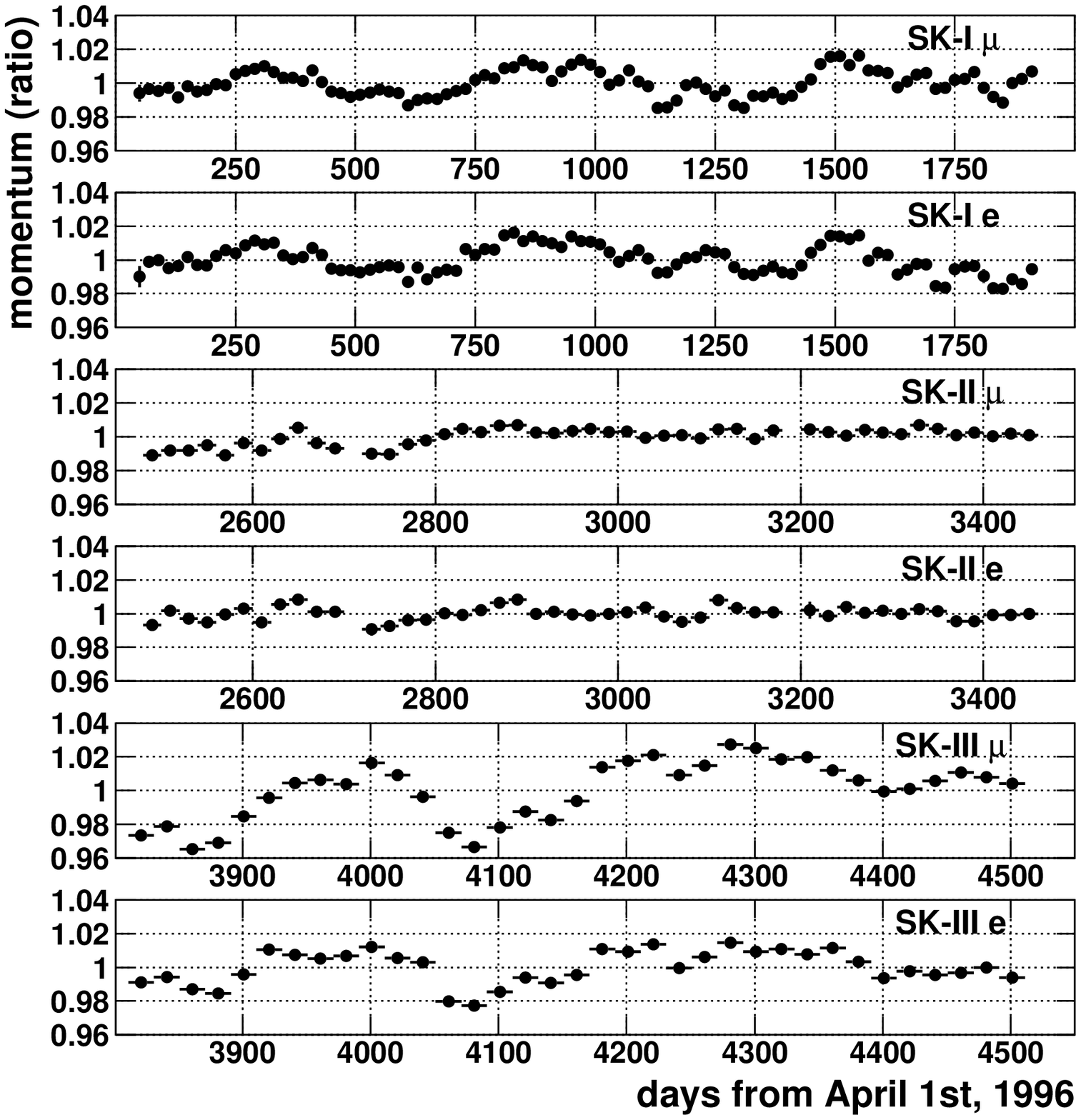}
  \caption
  {\protect \small 
The time variation of 
the reconstructed momentum/range of stopping muons
and the reconstructed momentum of decay electrons 
as a function of elapsed days
with respect to the mean value 
for SK-I, SK-II, and SK-III from top to bottom, respectively.
The largest variation (RMS) among these two calibration sources
during SK-I, SK-II, and SK-III was 0.88\%, 0.55\%, and 1.79\%, respectively.
The time variation was relatively larger in SK-III 
due to its poorer water quality. 
}
\label{timevari}
\end{center}
\end{figure}

The uncertainty in the detector's energy calibration (the energy scale)
is one of the most important for the proton decay searches.
It is estimated by systematic studies of the differences 
between the absolute momenta of data and MC in several 
control samples. 
At SK these control samples are chosen to probe 
a variety of energies where neutrino interactions and 
nucleon decay searches are relevant, among which 
are cosmic ray muons that stop in the detector, their 
associated decay electrons (Fig.~\ref{decay-e}), and $\pi^0$s 
produced in neutral current atmospheric neutrino interactions 
(Fig.~\ref{pi0mass}).
The uncertainty of the absolute energy scale is estimated
to be less than 0.74\%, 1.60\%, and 2.08\%
for SK-I, SK-II, and SK-III, respectively.
The time variation of 
the reconstructed momenta of these muon and electron samples
during each of the SK run periods (Fig.~\ref{timevari})
is also taken into account.
Overall, the uncertainty on the energy scale 
is 1.1\%, 1.7\%, and 2.7\% in SK-I, SK-II, and SK-III, respectively.
Further, the directional dependence of decay electron momenta 
is used to estimate the asymmetry of the energy scale.
During SK-I, SK-II, and SK-III, the asymmetry was estimated to be 
0.6\%, 0.6\%, and 1.3\% respectively.
These errors on the SK energy scale are incorporated 
into the proton decay analysis as discussed in Sec.~VI-B-2.

\subsection{$K^0_L$ event reconstruction}

The $K^0_L$ event reconstruction needs two vertices,
one for the $\mu^+$ candidate from the proton decay and 
another for particles produced in the delayed $K^0_L$ decay.
The difference between the two vertices is used to
distinguish the $K^0_L$ decay signal from the atmospheric neutrino background.
However, the standard reconstruction described cannot by itself 
reconstruct two vertices.
For instance, the standard ring counting algorithm uses a cut on 
the PMT hit timing after correcting for the time of flight 
taken from the initial vertex to help isolate genuine 
Cherenkov rings. Since the light from the particles emerging from 
the $K^{0}_L$ is both delayed in time and spatially separated from 
this initial vertex, the inclusion of this PMT timing cut degrades
the performance of the ring counting algorithm on these events.
Therefore, we developed a specialized reconstruction.
In the first step of the $K^{0}_L$ reconstruction, 
the standard reconstruction algorithm is applied without making 
any cut on the PMT hit timing during the ring counting stage. 
It outputs an 
initial vertex, the number of candidate rings, 
and a direction for each ring in much the same way 
as the standard reconstruction.

In the second step of the reconstruction~\cite{rec2}, 
an event's hit PMTs
are divided into two groups for each ring found by the first step. 
The first,
called the inner domain, is a collection of hit PMTs that fall within a
cone that is 10$^{\circ}$ larger than the ring's reconstructed Cherenkov
angle. The second group, called the outer domain, are the remaining
collection of hit PMTs that are not part of the inner domain. The charge
and time information of PMTs in the outer domain are zeroed and part of
the standard reconstruction is reapplied using only information from
PMTs in the inner domain. At this stage the vertex reconstruction, PID,
and momentum reconstruction are repeated, and the result is stored. A
precise vertex fit, which adjusts the initial candidate vertex along the
ring's reconstructed momentum using the expected charge distribution for
a particle with the ring's new PID,
is applied. 
The vertex resolution is about 0.3~m~\cite{skatmosc}.
Finally, an algorithm designed to
distinguish protons from muons~\cite{maxim}, 
is applied to provide an additional
PID handle. Once the $\mu^+$ candidate from the original proton's decay is
identified, this precision vertex will be identified as that of the
initial decay. The secondary vertex from the $K^0_L$ decay, is estimated by
zeroing the charge and time information of the inner domain PMTs and
applying the point fit algorithm to the outer domain. 
Since $\pi^{\pm}$ from the
$K^0_L$ decay may be just above Cherenkov threshold, 
these pions may not appear in
the detector as clear Cherenkov rings. Using the outer domain to perform
the secondary vertex fit in this way incorporates light from even these
particles to provide a better estimate of the point where the $K^0_L$ decays.
These reconstructions in the second step 
are repeated for each ring found by the first step.

\begin{figure}[htbp]
\begin{center}
  \includegraphics[width=8cm,clip]{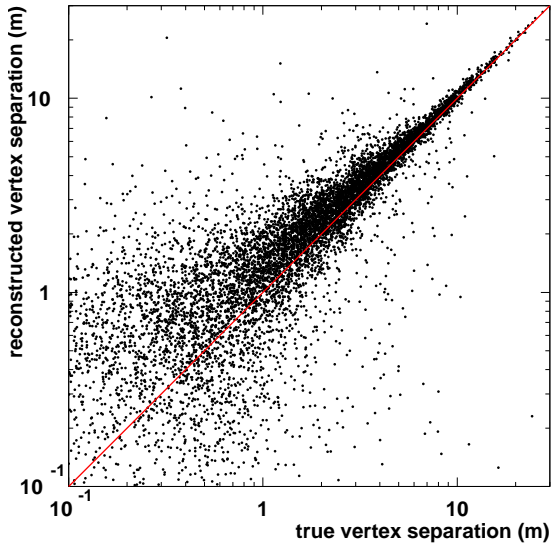}
  \caption
  {\protect \small 
Correlation between true and reconstructed vertex separations
for the $p\rightarrow\mu^+ K^0_L$ decay MC events in SK-I
after the event selections (D1-D5) 
described in Section~VI-A.}
  \label{vs1}
\end{center}
\end{figure}

\begin{figure}[htbp]
\begin{center}
  \includegraphics[width=8cm,clip]{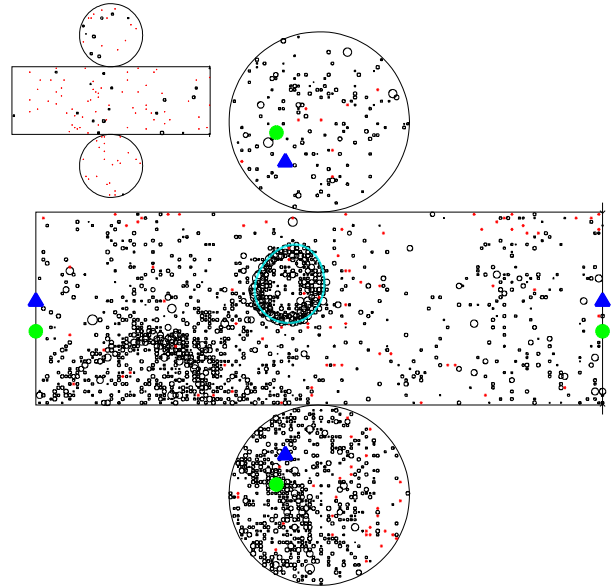}
  \caption
  {\protect \small 
Typical event display of $p\rightarrow\mu^+ K^0_L$,
$K^0_L\rightarrow\pi^{\mp}e^{\pm}\nu$ MC event in SK-I.
The upper (lower) visible ring are from true $\mu^+$ ($e$).
The cyan thick circle shows the $\mu^+$ candidate ring.
The blue triangle and green circle show the reconstructed vertex
position horizontally and vertically projected on the detector wall
for the $\mu^+$ and remaining particle candidates, respectively.
For this event, reconstructed and true vertex separations
are 6.58~m and 6.33~m, respectively.
}
  \label{klpdkmc_evdisp}
\end{center}
\end{figure}

For each event, the ring with $\mu$-like momentum 
closest to 326.5~MeV/$c$ is chosen as the $\mu^+$ candidate.
A vertex separation parameter is defined as the 
distance between the $\mu^{+}$ (primary) candidate vertex 
and the $K^0_L$ (secondary) vertex along the $\mu^{+}$ candidate's 
momentum vector. This parameter is used to define a cut  
to help reduce atmospheric neutrino backgrounds and is described below.
Figure~\ref{vs1} shows the typical correlation between the true vertex 
separation, extracted using MC information, 
and the reconstructed vertex separation
for $p\rightarrow\mu^+ K^0_L$ decay MC events. A clear 
correlation between the two can be seen.
A typical event display from one MC event showing both 
the primary and secondary fitted vertices appears in Fig.~\ref{klpdkmc_evdisp}.
In this case the reconstruction has correctly identified the 
$\mu^{+}$ (thick cyan ring) and the primary and secondary 
vertex are represented by the blue triangle and green circle, respectively.

\begin{figure}[htbp]
\begin{center}
  \includegraphics[width=9cm,clip]{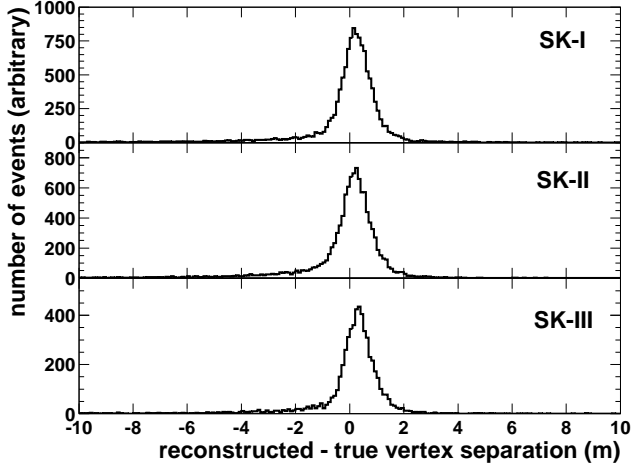}
  \caption
  {\protect \small 
Reconstructed - true vertex separation
for the $p\rightarrow\mu^+ K^0_L$ decay MC events
in SK-I, SK-II, and SK-III (from top to bottom), respectively,
after the event selections (D1-D5) described in Section~VI-A.}
  \label{rec-true_vs}
\end{center}
\end{figure}

The difference between the reconstructed and true 
vertex separation is shown in Fig.~\ref{rec-true_vs} for 
proton decay events in each of the SK run periods.
A slight bias in this distribution, represented by the shift 
of the peak away from zero, is used to estimate the uncertainty 
coming from the vertex separation cut discussed in Sec.~VI-B-2.
This shift is typically $O(10)~\mbox{cm}$ for both 
proton decay and atmospheric neutrino MC events. 
The vertex separation resolution for these two 
MC sets are about 0.8~m and 1.1~m, respectively. 
Due to the cleaner back-to-back event topology of the 
$K^0_L$ decay events, the separation resolution 
for the proton decay events is slightly better than that
for the atmospheric neutrino events.
Most of this resolution comes from the application 
of the point fit algorithm in determining the secondary 
vertex.  

\section{DATA ANALYSIS}

\subsection{Event Selection}

The search for $p\rightarrow\mu^+K^0$ decay is performed 
by searching for decays into $\mu^+K^0_S$ and $\mu^+K^0_L$
separately. The FCFV events (Sec.~IV) 
are first passed through the $K^{0}_S$ selection
and failing events at the $K^0_S$ selection 
are then additionally subjected to the $K^{0}_L$ search. 
The $K^{0}_S$ search criterion are not changed 
with respect to the previous analysis \cite{kk}, 
but are reviewed here for completeness. 

To maximize the analysis' sensitivity to the $K^{0}_S$ mode, 
three distinct selections designed to extract events corresponding to 
each of the $K^{0}_S$ decay's final states are used. 
It should be noted that by construction there is no kinematic overlap among 
the methods.
For $K^0_S\rightarrow\pi^0\pi^0$ decays (30.7\% branching fraction),
the proton decay selection criteria are as follows:
\begin{itemize}
\item (A1) 3-5 rings, corresponding to the initial $\mu^{+}$ and 
           the gammas emerging from the $\pi^{0} $ decays
\item (A2) one $\mu$-like ring, to be identified as the $\mu^{+}$ candidate,
           and the other rings in the event are $e$-like
\item (A3) one Michel electron, from the decay of the $\mu^{+}$
\item (A4) 150~MeV/$c$ $<$ $p_{\mu}$ $<$ 400~MeV/$c$
\item (A5) 400~MeV/$c$$^2$ $<$ $m_{K^0}$ $<$ 600~MeV/$c$$^2$
\item (A6) $p_p$ $<$ 300~MeV/$c$
\item (A7) 750~MeV/$c$$^2$ $<$ $m_p$ $<$ 1000~MeV/$c$$^2$
\end{itemize}

\begin{figure*}[htbp]
\begin{center}
  \includegraphics[width=\linewidth]{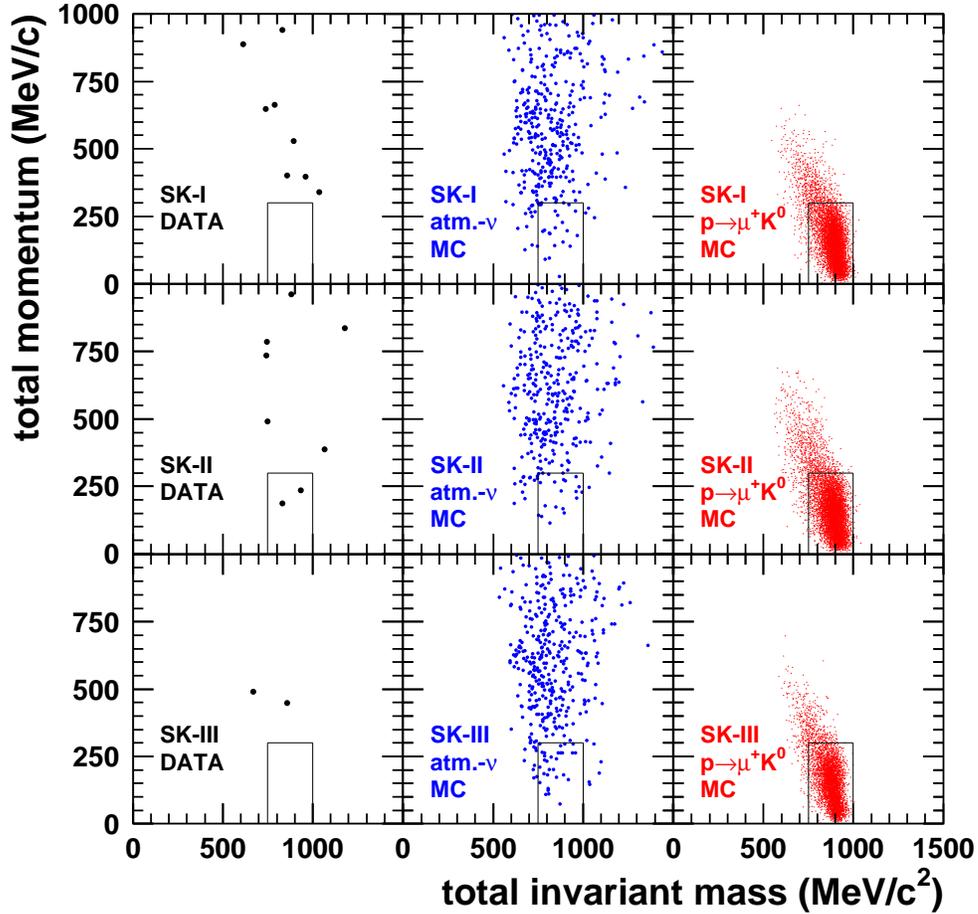}
  \caption
  {\protect \small 
Total invariant mass and momentum
for events that satisfy criteria (A1-A5)
in SK-I, SK-II, and SK-III
from top to bottom.
The box shows criteria (A6, A7).
From left to right,
data (1489~days, 799~days, 518~days),
atmospheric neutrino MC (200~years, 172~years, 194~years),
and $p\rightarrow\mu^+K^0$ MC are shown.}
  \label{pi0_2d}
\end{center}
\end{figure*}

\noindent Here $p_{\mu}$ refers to the $\mu^+$ candidate momentum,
$m_{K^0}$ is the reconstructed ${K^0}$ invariant mass
using only the $e$-like rings, 
and $p_p$ ($m_p$)
is the total momentum (invariant mass)
using all rings in the event. 
Figure~\ref{pi0_2d} shows the $m_p$ and $p_p$ distributions
after applying these criteria (A1-A5).
Under the hypothesis of proton decay,
initial proton momenta of up to 300~MeV/$c$ are 
considered in the search because the Fermi momentum 
of protons in $^{16}$O nuclei may reach about 200~MeV/$c$.
This, in addition to the possibility of missing gammas 
from the $\pi^{0}$ decay, is similarly the reason 
for the small lower bound on the $m_p$ cut.

After this event selection the remaining background consists
primarily of events from $\nu_e$ or $\nu_{\mu}$ 
charged current (CC) multiple pion production interactions,
and neutral current (NC) deep inelastic scattering processes,
from the atmospheric neutrino MC.
These backgrounds often have a single $\mu$-like ring back-to-back
with several $e$-like rings and are consistent with the 
expected event signature of the signal. 

\begin{figure*}[htbp]
\begin{center}
  \includegraphics[width=\linewidth]{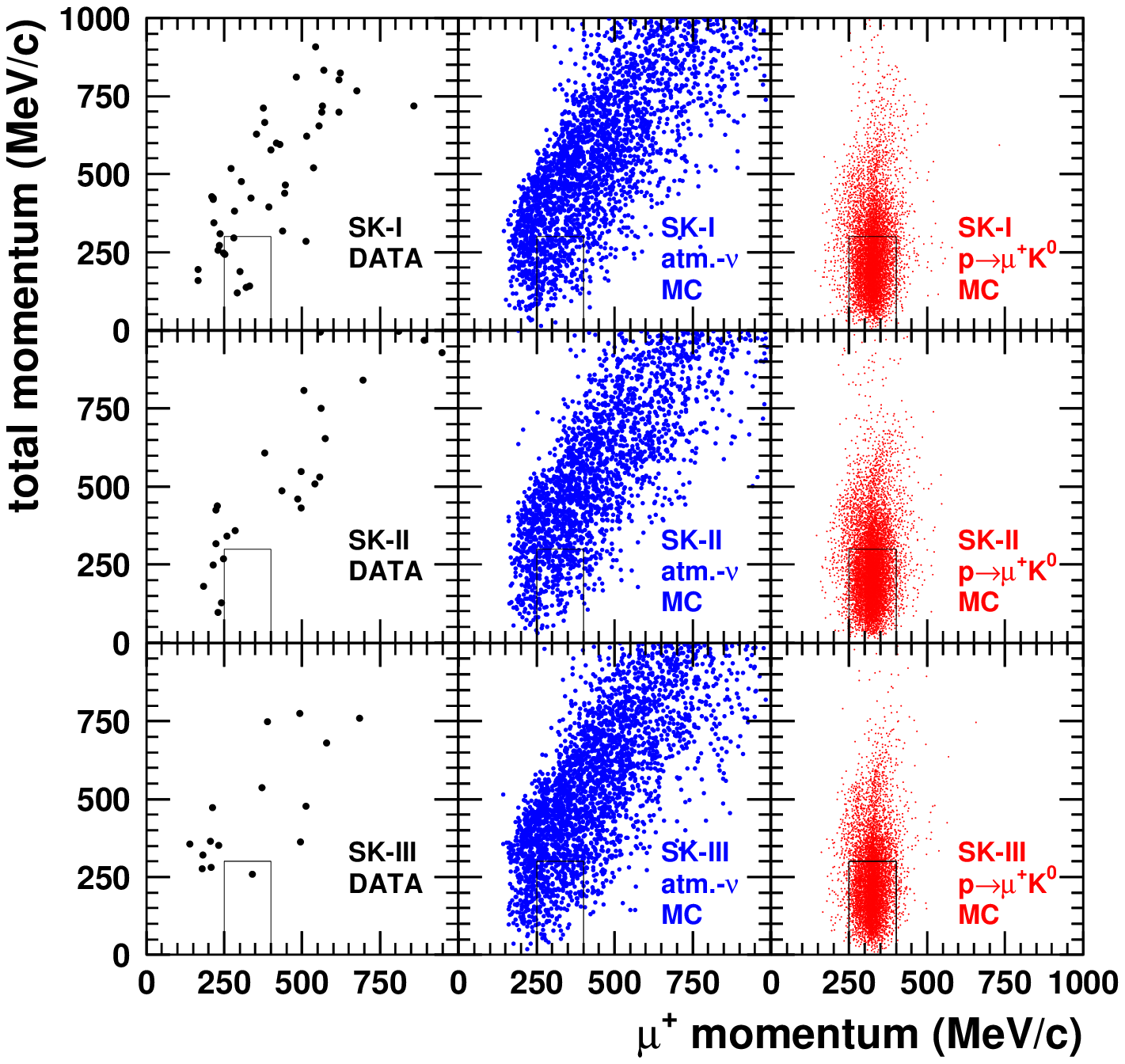}
  \caption
  {\protect \small
$\mu^+$ momentum and total momentum
for events that satisfy criteria (B1-B3)
in SK-I, SK-II, and SK-III
from top to bottom.
The box shows criteria (B4, B5).
From left to right,  
data (1489~days, 799~days, 518~days),
atmospheric neutrino MC (200~years, 172~years, 194~years),
and $p\rightarrow\mu^+K^0$ MC are shown.}
  \label{pi1_2d}
\end{center}
\end{figure*}

\begin{figure*}[htbp]
\begin{center}
  \includegraphics[width=\linewidth]{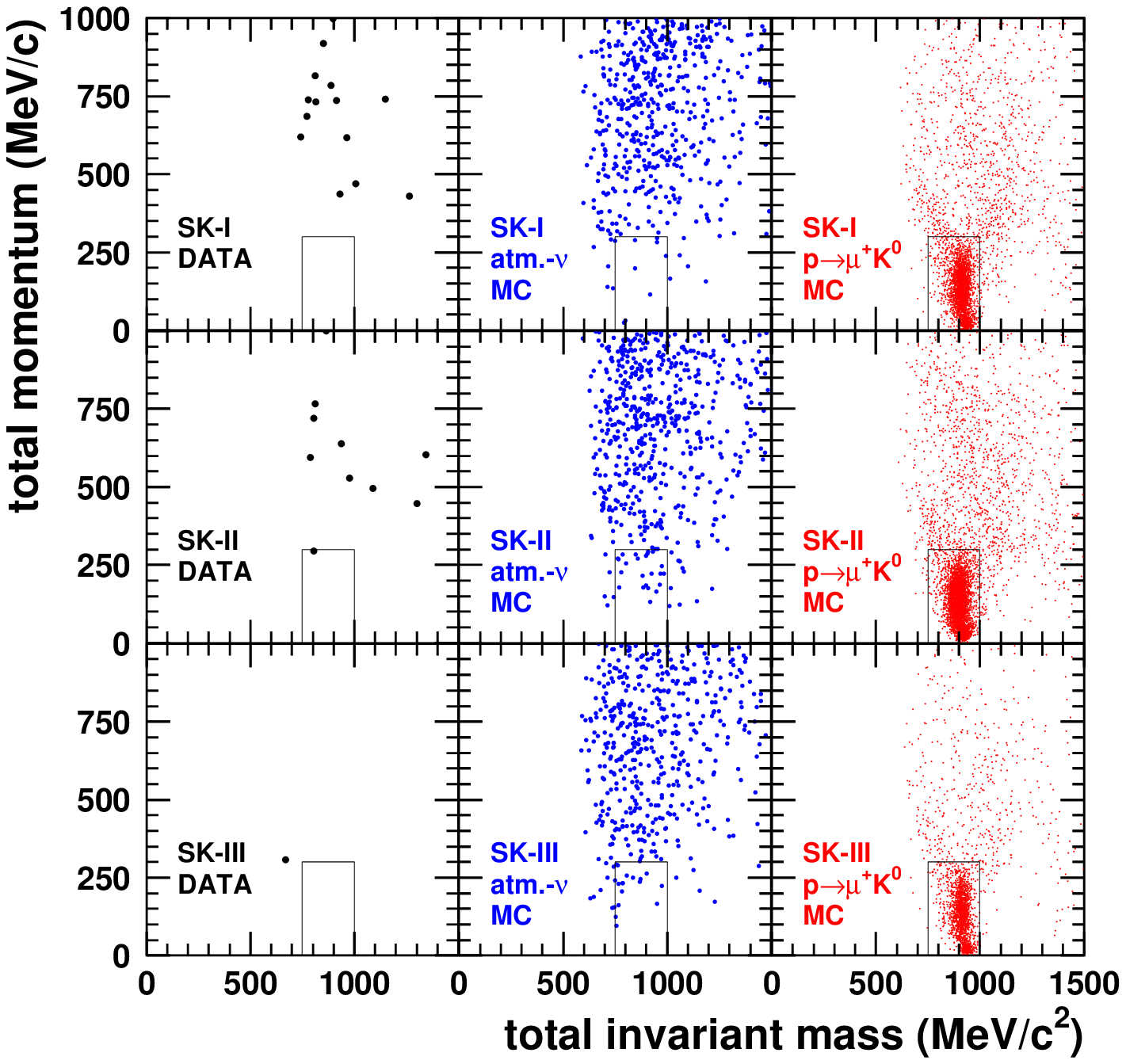}
  \caption
  {\protect \small
Total invariant mass and total momentum
for events that satisfy criteria (C1-C3)
in SK-I, SK-II, and SK-III
from top to bottom.
The box shows criteria (C4, C5).
From left to right,
data (1489~days, 799~days, 518~days),
atmospheric neutrino MC (200~years, 172~years, 194~years),
and $p\rightarrow\mu^+K^0$ MC are shown.}
  \label{pi2_2d}
\end{center}
\end{figure*}

To search for proton decays in which the $K^0_S$ decays into $\pi^{+}\pi^{-}$
(69.2\% branching fraction) two different methods are used.
The first method (Method 1) is designed to search for 
events where one of the outgoing pions does not produce 
a clear Cherenkov ring, 
for instance when its momentum is below the Cherenkov production threshold, 
or when it captures on a $^{16}$O nucleus.
The search criteria are:
\begin{itemize}
\item (B1) two rings
\item (B2) both rings are $\mu$-like, one from the $\mu^{+}$
           and another from one of the charged pions
\item (B3) two Michel electrons, from the decays of the 
           primary $\mu^{+}$ and that from the charged pion
\item (B4) 250~MeV/$c$ $<$ $p_{\mu}$ $<$ 400~MeV/$c$
\item (B5) $p_p$ $<$ 300~MeV/$c$
\end{itemize}
In this method the most energetic $\mu$-like ring 
is taken to be the $\mu^{+}$ emerging from the initial 
proton decay and is assigned a $\mu$-like momentum. 
On the other hand, the second energetic ring is assumed to be the
charged pion and is assigned a $\pi$-like momentum. 
Figure~\ref{pi1_2d} shows the $p_{\mu}$ and $p_p$ distributions
after applying these criteria (B1-B3).
The second method (Method 2) searches for proton decays 
with $K^{0}_S \rightarrow \pi^{+}\pi^{-}$ 
in which all of the emerging charged particles produce Cherenkov radiation in the detector.
In this case the selection cuts become:
\begin{itemize}
\item (C1) three rings
\item (C2) one or two Michel electrons
\item (C3) 450~MeV/$c$$^2$ $<$ $m_{K^0}$ $<$ 550~MeV/$c$$^2$
\item (C4) $p_p$ $<$ 300~MeV/$c$
\item (C5) 750~MeV/$c$$^2$ $<$ $m_p$ $<$ 1000~MeV/$c$$^2$
\end{itemize}
Figure~\ref{pi2_2d} shows the $m_p$ and $p_p$ distributions
that result from applying the criteria (C1-C3).
The primary background to this search arises from $\nu_{\mu}$ CC 
interactions with single- or multiple-pions produced in the final state. 

\begin{figure}[htbp]
\begin{center}
  \includegraphics[width=9cm,clip]{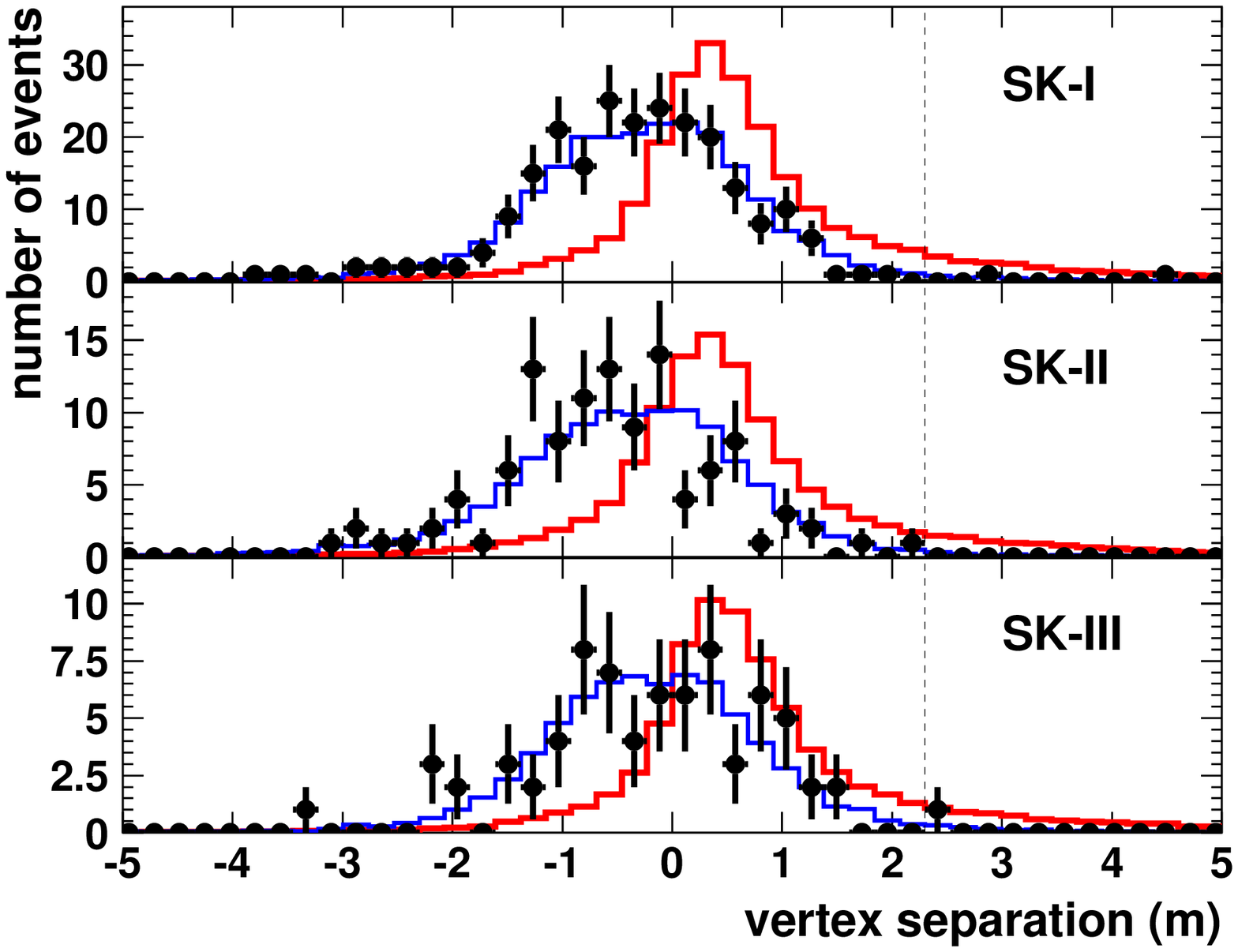}
  \caption
  {\protect \small
Vertex separation distributions 
for events that satisfy criteria (D1-D5)
in SK-I, SK-II, and SK-III from top to bottom.
Positive values  of the vertex separation indicate that
the point fit vertex is separated from
the $\mu^+$ vertex in the direction opposite 
of the $\mu^+$.
Data (black dot),
atmospheric neutrino MC (blue thinner histogram),
and $p\rightarrow\mu^+K^0$ MC (red thicker histogram) are shown.
The MC samples are normalized to data by entries.
The cut value corresponding to (D6) is shown as broken line.}
  \label{vs2}
\end{center}
\end{figure}

To search for proton decays with a $K^0_L$ decay
the following criteria are applied:
\begin{itemize}
\item (D1) 500(100)~pe $<$ total charge $<$ 8000(3000)~pe
in SK-I,III(SK-II),
\item (D2) the number of rings without the PMT hit-timing cut $>$ 1
\item (D3) at least one $\mu$-like ring
that is neither $e$-like nor proton-like
\item (D4) the number of Michel electrons $>$ 0
\item (D5) 260~MeV/$c$ $<$ $p_{\mu}$ $<$ 410~MeV/$c$
\item (D6) vertex separation between $\mu^{+}$ candidate 
           and $K^{0}_L$ decay is greater than 2.3~m
\end{itemize}
The $\mu^+$-candidate is taken to be the $\mu$-like ring
with momentum closest to 326.5~MeV/$c$.
At the third cut (D3), the  following criteria are required
to increase purity of the $\mu$-like rings.
In addition to requiring a $\mu$-like classification by the 
standard PID, the result of the proton ID algorithm must also be 
$\mu$-like. Further, the Cherenkov opening angle 
is required to be greater than 30$^{\circ}$,
The cut value at (D6) has been chosen to 
maximize this analysis' sensitivity to the proton partial lifetime 
based on studies of the proton decay and atmospheric neutrino MC.
Figure~\ref{vs2} shows the vertex separation distribution
after applying criteria (D1-D5).
The net decay length of the $p\rightarrow\mu^+K^0_L$ MC is 
about 2~m because roughly 70\% of the $K^0_L$ undergo hadronic interactions in the water 
before decaying. 
There is good agreement between the data and 
the atmospheric neutrino MC.

\begin{figure}[htbp]
\begin{center}
  \includegraphics[width=8cm,clip]{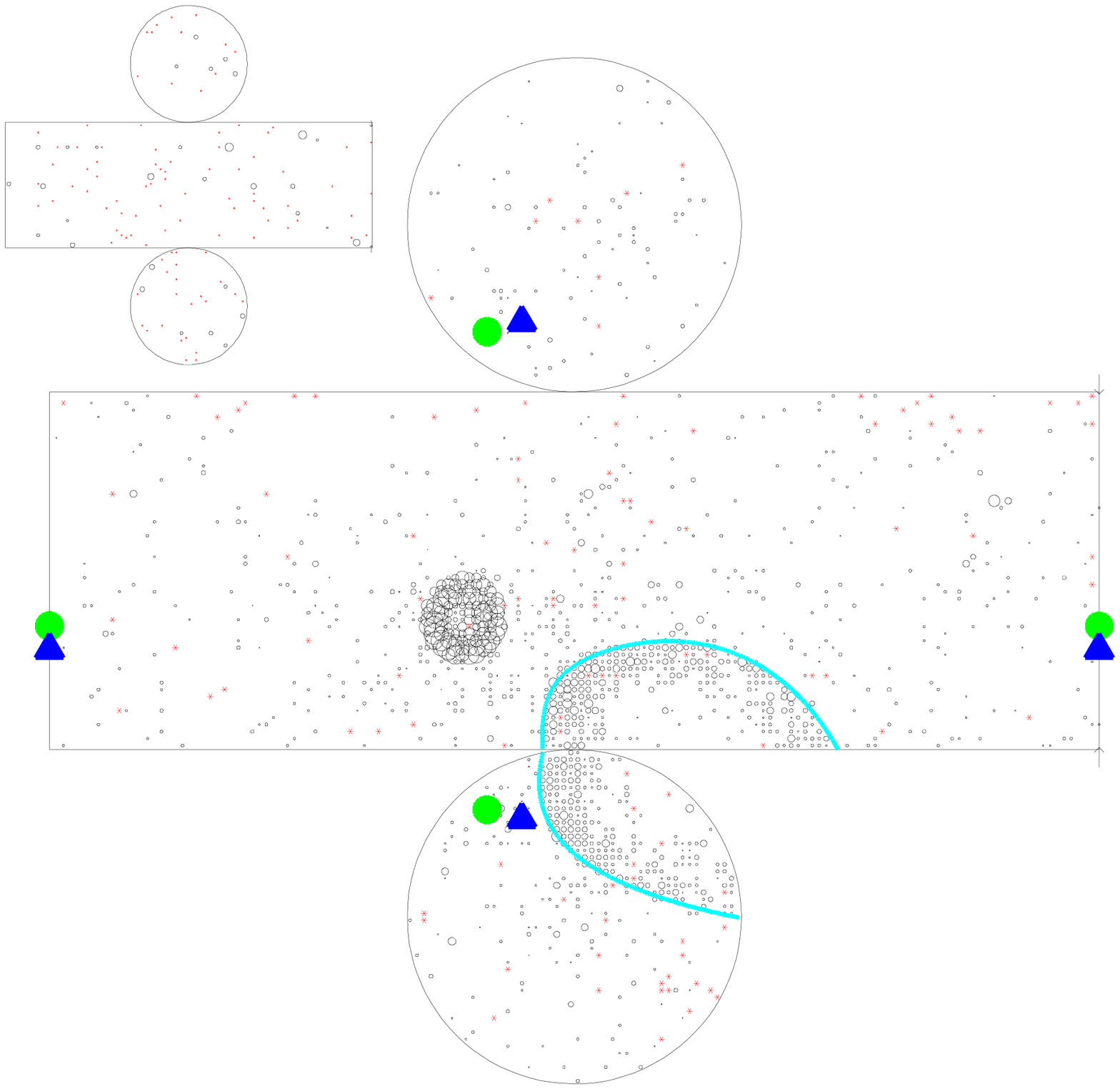}
  \caption
  {\protect \small 
Typical event display of the remaining atmospheric
neutrino background MC event in the $K^0_L$ search in SK-I.
This is from $\nu_{\mu}$ charged current quasi-elastic interaction.
The right (left) visible rings are from
true proton (muon).
The cyan ring shows the $\mu^+$ candidate reconstructed ring.
The blue triangle and green circle show the reconstructed vertex 
position horizontally and vertically projected on the detector wall
for the $\mu^+$ and remaining particle candidates, respectively.
}
  \label{klbgmc_evdisp}
\end{center}
\end{figure}

The dominant background sources after the event selection are 
primarily $\nu_{\mu}$ CC quasi-elastic and single-pion production interactions.
In these backgrounds
the recoiling proton from the neutrino interaction 
is above Cherenkov threshold and is selected as the $\mu^+$-candidate.
For this reason the primary vertex can be pulled 
along the direction of the $\mu^+$ candidate momentum, increasing 
the separation between the two reconstructed vertices. 
Figure~\ref{klbgmc_evdisp} shows a typical event display
of these backgrounds. In this particular event 
the reconstructed vertex separation is 3.9~m.
%
%
%

\subsection{Result}

The signal detection efficiency,
the expected atmospheric neutrino background rate,
and the number of candidate events for each of the search methods 
are summarized in Table~\ref{result_table}.
\begin{table*}[htbp]
\begin{center}
    \begin{tabular}{|l|l|c|c|c|c|}
      \hline \hline
detector (exposure) & search method & efficiency & background & candidate & lower limit \\
         &               &  ($\%$)    &            &          & ($\times$10$^{33}$~years) \\
      \hline \hline
SK-I (91.7~kton$\cdot$years) & $K^0_S\rightarrow\pi^0\pi^0$ & 7.0$\pm$0.5 & 0.37$\pm$0.05 (4.0$\pm$0.5)   & 0 & 0.92\\
     & $K^0_S\rightarrow\pi^+\pi^-$ Method~1 & 10.6$\pm$1.0 & 3.0$\pm$0.5 (33$\pm$5) & 6 & 0.42\\
     & $K^0_S\rightarrow\pi^+\pi^-$ Method~2 & 2.5$\pm$0.2 & 0.12$\pm$0.08 (1.3$\pm$0.9) & 0 & 0.32\\
     & $K^0_L$                      & 3.8$\pm$0.7 & 3.5$\pm$1.1 (38$\pm$12) & 2 & 0.30\\
\hline
SK-II (49.2~kton$\cdot$years) & $K^0_S\rightarrow\pi^0\pi^0$ & 6.2$\pm$0.7 & 0.20$\pm$0.05 (4.1$\pm$1.0) & 2 & 0.19\\
     & $K^0_S\rightarrow\pi^+\pi^-$ Method~1 & 10.3$\pm$1.2 & 1.6$\pm$0.4 (33$\pm$8) & 0 & 0.71\\
     & $K^0_S\rightarrow\pi^+\pi^-$ Method~2 & 2.4$\pm$0.2 & 0.23$\pm$0.08 (4.7$\pm$1.6) & 1 & 0.11\\
     & $K^0_L$                      & 3.3$\pm$0.6 & 1.4$\pm$0.5 (29$\pm$10) & 0 & 0.21\\
\hline
SK-III (31.9~kton$\cdot$years) & $K^0_S\rightarrow\pi^0\pi^0$ & 6.7$\pm$0.7 & 0.19$\pm$0.04 (6.0$\pm$1.3) & 0 & 0.30\\
     & $K^0_S\rightarrow\pi^+\pi^-$ Method~1 & 10.3$\pm$1.8 & 1.2$\pm$0.2 (38$\pm$6) & 1 & 0.32\\
     & $K^0_S\rightarrow\pi^+\pi^-$ Method~2 & 3.0$\pm$0.2 & 0.09$\pm$0.02 (2.8$\pm$0.6) & 0 & 0.14\\
     & $K^0_L$                      & 3.8$\pm$0.7 & 1.3$\pm$0.6 (41$\pm$19) & 1 & 0.11\\
      \hline
SK-I+SK-II+SK-III (172.8~kton$\cdot$years) & combined & & & & 1.6\\
      \hline \hline
    \end{tabular}
  \caption{\protect \small
Summary of the $p\rightarrow\mu^+ K^0$ search.
Systematic errors are shown in the signal efficiencies and background rates
(see Table~\ref{syst_table} for more detail).
The numbers in the parentheses of the background columns 
show the expected background rates in Megaton$\cdot$years$^{-1}$.}
  \label{result_table}
\end{center}
\end{table*}

The detection efficiencies in SK-II are lower than in SK-I and SK-III
and their differences are at most 20\% as shown in Table~I.
The estimated background rates of SK-I, SK-II, and SK-III
are on the order of a few events
and the dominant source of uncertainty
is the MC statistical error as described in Sec.~VI-B-2.
For instance, the statistical error on the background
in the $K^0_S\rightarrow\pi^+\pi^-$ Method~2 search is 20-40\%.

\begin{figure*}[htbp]
\begin{center}
\includegraphics[width=14cm,clip]{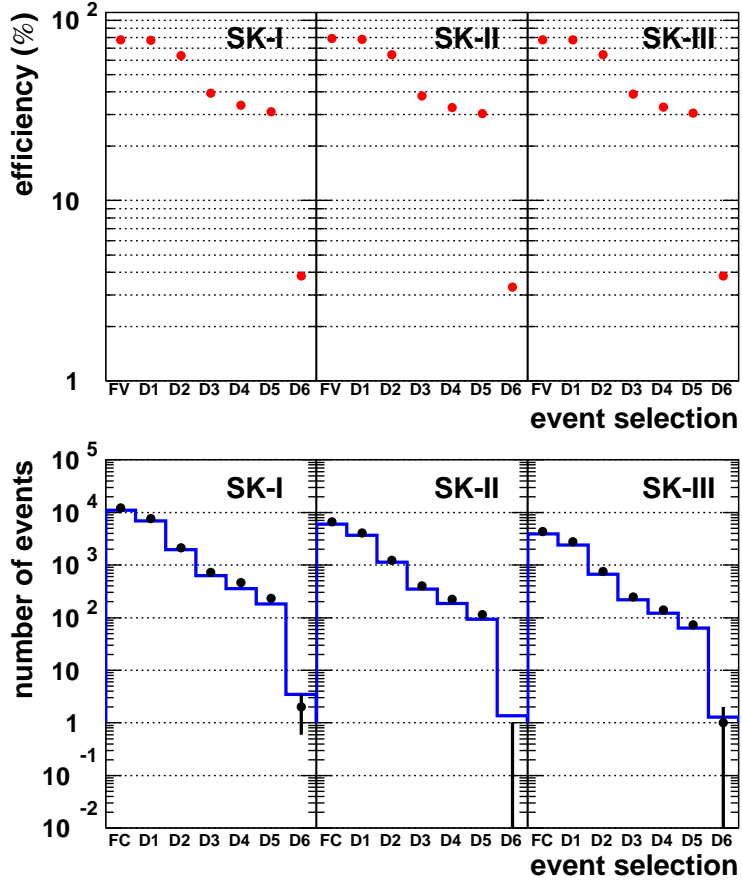}
\caption{\protect \small
The  $K^0_L$ search;
(upper plot)
the signal efficiencies along event selections are shown
in SK-I, SK-II, and SK-III from left to right, respectively
(``FV'' stands for the number of $p\rightarrow\mu^+K^0$ MC events
generated in true fiducial volume (22.5~kton).
The events selected by the $K^0_S$ searches are excluded), 
(lower plot)
the number of data candidates (black circle) 
and the expected atmospheric neutrino background rate (blue histogram) 
along event selections are shown
in SK-I, SK-II, and SK-III from left to right, respectively
(``FC'' stands for the FCFV events described in Sec.~IV).
Only statistical errors are shown.
There are no events remaining in the SK-II data after selection (D6).}
\label{cutrate_kl}
\end{center}
\end{figure*}

The number of data and the expected background 
and signal efficiencies at each step of the event selection are compared
for all four searches. 
For example,
Figures~\ref{cutrate_kl} shows the $K^0_L$ search.
After each event selection, the number of remaining events 
in the data is consistent with the background expectation in each 
of the SK run periods.
The Poisson probability to observe the number of candidates
after all event selections for each search was calculated. 
For example, the only probability smaller than 5\% 
is seen in the SK-II $K^0_s\rightarrow\pi^0\pi^0$ search (2\%).
However, the number of candidates is smaller than 
the expected backgrounds in SK-I and SK-III for this search.
The sums of the expected backgrounds and the number of candidates
from all SK run periods for the $K^0_S$ and $K^0_L$ searches  
are 13.2 and 13, respectively.
They are consistent and  therefore we conclude that there 
is no evidence for proton decay found in this analysis.

\begin{figure}[htbp]
\begin{center}
  \includegraphics[width=8cm,clip]{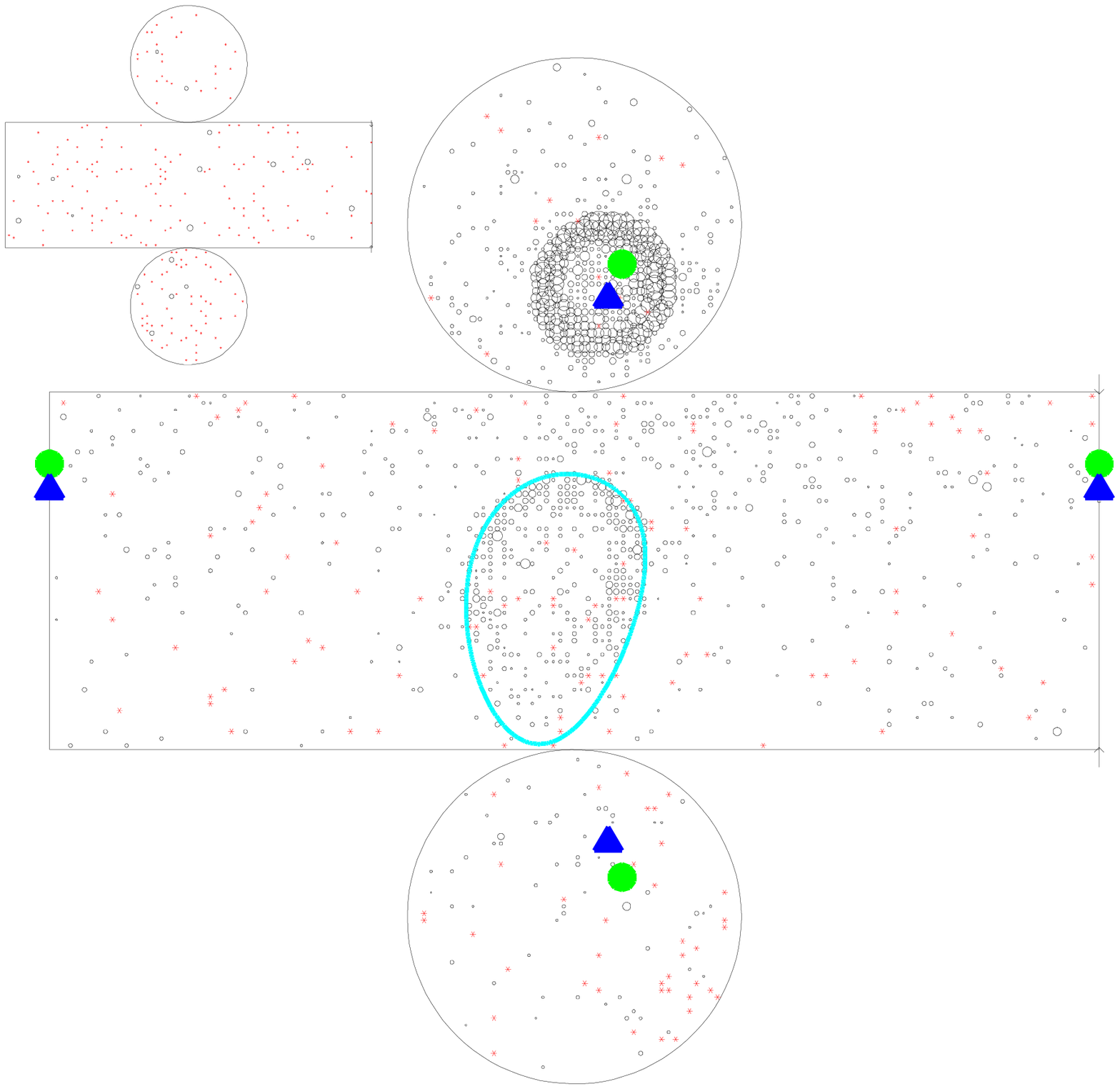}
  \caption
  {\protect \small 
Typical event display of the real data candidate event
in the $K^0_L$ search in SK-I.
The cyan ring shows the reconstructed ring edge
of the $\mu^+$ candidate ring.
The blue triangle and green circle show the reconstructed vertex
position horizontally and vertically projected on the detector wall
for the $\mu^+$ and remaining particle candidates, respectively.
The reconstructed vertex separation for this event is 4.6~m.
The reconstructed Cherenkov angle is 30.6$^{\circ}$
while the expectation from the $\mu$-like momentum (291~MeV/$c$)
is 37.6$^{\circ}$.
}
  \label{klcand_evdisp}
\end{center}
\end{figure}

All the candidate events in Table~\ref{result_table}
were inspected by hand.
By checking event display with the reconstructed information,
all the events show Sub-GeV (visible energy is less than 1330~MeV)
back-to-back multi visible rings and
no obvious mis-reconstruction.
For example, 
Figure~\ref{klcand_evdisp} shows 
typical real data candidate event in the $K^0_L$ search.
This event looks consistent with the $\nu_{\mu}$ background
where the visible proton is identified as the $\mu^+$ candidate ring
(see Figure~\ref{klbgmc_evdisp}).

\subsubsection{Lifetime Limit}

In the absence of significant excess of observed data
above the background expectation
we calculate a lower limit on the proton partial lifetime
using a Bayesian method~\cite{bayesian}. 
The calculation used here follows the one from the previous analysis \cite{kk}.

The decay of the proton is assumed to follow a Poisson probability, 
$P(\Gamma|n_i)$, and is expressed as:
\begin{eqnarray}
P(\Gamma|n_i) = \int\int\int\frac{e^{-(\Gamma\lambda_i\epsilon_i+b_i)}
(\Gamma\lambda_i\epsilon_i+b_i)^{n_i}}{n_i!}\nonumber\\
P(\Gamma)P(\lambda_i)P(\epsilon_i)P(b_i)d\lambda_id\epsilon_idb_i\nonumber
\end{eqnarray}
where $n_i$ is the number of candidate events
in the $i$-th proton decay search 
($i$ = 1, 2, 3, and 4 for $K^0_S\rightarrow\pi^0\pi^0$,
$K^0_S\rightarrow\pi^+\pi^-$ Method~1,
$K^0_S\rightarrow\pi^+\pi^-$ Method~2,
and $K^0_L$ searches, respectively, in SK-I,
$i$ = 5-8 in SK-II,
and $i$ = 9-12 in SK-III
in Table~\ref{result_table}),
$\Gamma$ is the total decay rate,
$\lambda_i$ is the detector exposure,
$\epsilon_i$ is the detection efficiency 
including the meson branching ratio, and $b_i$ is the number of expected background events.
The probability density function, $P(\Gamma)$, 
describing the prior expectation of decay rate 
is taken to be $P(\Gamma) = 1$ for $\Gamma > 0$ and zero otherwise.  

Introduction of systematic error effects is done through prior probabilities 
for the detector exposure $P(\lambda_i)$, efficiency $P(\epsilon_i)$,
 and background expectation $P(b_i)$. 
They are all assumed to be truncated Gaussian distributions of the form,
  \begin{eqnarray}
   & P(x_i) \propto
    \left\{
     \begin{array}{ll}
      \exp\left(-\frac{(x_i-x_{0,i})^2}{2\sigma_{x_i}^2}\right) 
& \left(x_i>0\right)\nonumber\\
      0 &\left(x_i \leq 0\right)
     \end{array}
    \right.\nonumber\\
   &(x_i = \lambda_i, \epsilon_i, b_i)\nonumber
  \end{eqnarray}
\noindent where $\lambda_{0,i}$ ($\sigma_{\lambda_i}$), 
$\epsilon_{0,i}$ ($\sigma_{\epsilon_i}$) 
and $b_{0,i}$ ($\sigma_{b_i}$) 
are the estimates (systematic errors) 
of the exposure, the detection efficiency, and the background, respectively.
Here $\epsilon_{0,i}$ and $b_{0,i}$ correspond to 
the $i$-th searches efficiency and background contamination, respectively.
These values are summarized in Table~\ref{result_table}.
The parameters $\sigma_{\epsilon_i}$ and $\sigma_{b_i}$
correspond to the total systematic error on the signal and 
background composition for the $i$-th search.
Systematic error estimations are discussed in the next 
section and Table~\ref{syst_table}
summarizes the errors for each run period and search mode. 
Finally, $\lambda_{0,i}$ represents the detector exposure,
taken to be product of the number of target protons in the 
detector and the livetime of each SK run. 
The corresponding uncertainty on the exposure, $\sigma_{\lambda_i}$, is 
assumed to be 1\%.

We calculate the lower limit of the nucleon decay rate,
$\Gamma_{\mathrm{limit}}$, using a 90\% confidence level (CL) as:
\begin{eqnarray}
\mathrm{CL}=\frac{\int^{\Gamma_{\mathrm{limit}}}_{\Gamma=0}\prod^N_{i=1}P(\Gamma|n_i)d\Gamma} {\int^\infty_{\Gamma=0}\prod^N_{i=1}P(\Gamma|n_i)d\Gamma}\nonumber
\end{eqnarray}
where $N$ (= 12) is the number of searches.
The lower lifetime limit:
\begin{eqnarray}
\tau/\mathrm{B}_{p\rightarrow\mu^+K^0}=\frac{1}{\Gamma_{\mathrm{limit}}}\nonumber
\end{eqnarray}
is set to be 1.6$\times$10$^{33}$ years at 90\% CL.
The result of the limit calculation
for each search method and data taking period
is shown in Table~\ref{result_table}.

\subsubsection{Systematic Error}

Table~\ref{syst_table}
summarizes the systematic errors on the
$K^0_S\rightarrow\pi^0\pi^0$,
$K^0_S\rightarrow\pi^+\pi^-$ Method~1,
$K^0_S\rightarrow\pi^+\pi^-$ Method~2,
and $K^0_L$ searches, respectively.

\begin{table*}[htbp]
\begin{center}
    \begin{tabular}{|l|l|c|c|}
      \hline \hline
search method & detector & efficiency ($\%$) & background ($\%$) \\
& & total (physics, detector) & total (physics, detector) \\
      \hline \hline
$K^0_S\rightarrow\pi^0\pi^0$ & SK-I & 7.5 (6.8, 3.2) & 14.2 (11.0, 9.2) \\
& SK-II & 10.8 (6.8, 8.4) & 23.4 (11.5, 20.4) \\
& SK-III & 10.0 (6.8, 7.4) & 19.1 (9.5, 16.5) \\
\hline
$K^0_S\rightarrow\pi^+\pi^-$ Method~1 & SK-I & 9.8 (8.8, 4.2) & 15.8 (15.2, 4.3) \\
& SK-II & 11.6 (8.8, 7.6) & 22.8 (16.5, 15.8) \\
& SK-III & 17.7 (8.9, 15.3) & 20.6 (16.2, 12.7) \\
\hline
$K^0_S\rightarrow\pi^+\pi^-$ Method~2 & SK-I & 7.5 (6.9, 2.8) & 63.7 (19.8, 60.5) \\
& SK-II & 7.1 (6.9, 1.8) & 34.4 (12.7, 32.0) \\
& SK-III & 7.9 (7.2, 3.4) & 24.6 (11.6, 21.7) \\
\hline
$K^0_L$ & SK-I & 17.2 (9.6, 14.3) & 31.9 (9.9, 30.3) \\
& SK-II & 19.2 (9.7, 16.6) & 38.9 (9.9, 37.6) \\
& SK-III & 18.4 (9.4, 15.8) & 48.4 (9.4, 47.5) \\
      \hline \hline
    \end{tabular}
  \caption{\protect \small
Summary of systematic errors.
The total systematic errors on the efficiency and background
are shown in the third and fourth columns, respectively.
The ``physics'' and ``detector'' in the parentheses represent
the systematic errors from uncertainty of the physics simulations
and the SK detector, respectively.}
  \label{syst_table}
\end{center}
\end{table*}

As for the physics simulation,
following uncertainties are taken into account.
In the signal efficiency,
differences in the Fermi momentum modeling~\cite{fermi,neut}
and uncertainties in the correlated decay probability~\cite{kk},
the $K^0_L\rightarrow K^0_S$ regeneration probability,
as well as the charged pion-nucleon cross section in water~\cite{kk}
are considered.
The dominant systematic error sources are correlated decay
and the pion-nucleon cross section. 
For the $K^0_L\rightarrow K^0_S$ regeneration,
the signal efficiencies for the default probability (about 0.1\%)
and about 0.2\% or 0\% 
(corresponding to $\pm$100\% error on the default probability)
are compared
and difference of the efficiencies is used
as the systematic error (less than 1\% for all the searches).
Total physics simulation errors are several percent
in all four searches in SK-I, SK-II, and SK-III.
In the background rate,
major systematic error sources
used in the SK atmospheric neutrino data analyses~\cite{skatmosc};
neutrino flux, neutrino cross sections,
and pion-nucleon interaction in oxygen, 
are taken into account
and the total simulation errors are about 10-20\%.

As for the SK detector,
uncertainty of the event selection cut value at each event selection step
is considered as well as the energy scale uncertainty.
Any detector uncertainty such as
light attenuation and scattering in water
is taken into account in the systematic error estimations
which compare the cut efficiencies between
control sample data and MC.
In the signal efficiency,
the total detector errors are several percent 
in the $K^0_S$ searches.
In the $K^0_S\rightarrow\pi^0\pi^0$ and
$K^0_S\rightarrow\pi^+\pi^-$ Method~1 searches,
the errors in SK-III are larger than those of SK-I by a factor of 2$\sim$3.
The larger errors in SK-III are due to its poorer water quality
which resulted in larger energy scale and ring counting systematics.
The total detector error in the $K^0_L$ search (a few 10\%)
is mostly due to the vertex separation.
The vertex separation error is estimated
by changing the cut value
corresponding to the difference
between true and reconstruction
described in Sec.~V-A.
In the background rates,
the dominant error (about 20-60\%)
comes from statistics of
the atmospheric neutrino events after all the event selections
in the $K^0_S\rightarrow\pi^+\pi^-$ Method~2 search.
The other dominant error (about 20-40\%) is
from the vertex separation in the $K^0_L$ search.

\subsection{Comparison with Previous Result}

In this section we describe the differences between the 
present analysis and the previous result~\cite{kk}. 
For the estimation of the signal efficiency 
the previous analysis only considered the $p\rightarrow\mu^+ K^0_S$ 
channel while the present work accounts for the entire $p\rightarrow\mu^+ K^0$ 
channel including the $K^{0}_L$ contribution. 
This contribution coupled with improvements to the event reconstruction
has resulted in an efficiency increase from 5.4\% to 7.0\% 
in the $K^0_S\rightarrow\pi^0\pi^0$ search
and from 7.0\% to 10.6\%
in the $K^0_S\rightarrow\pi^+\pi^-$ Method~1 search.
Signal efficiencies and the background rates for the other searches 
are consistent within the systematic errors of the two analyses.

Similar systematic error sources are incorporated in both analyses. 
For instance, the error on the signal efficiencies is about 10\%
in both. However, the systematic error on the background rate 
in this analysis, ranging from 20\% to 60\% across the searches, 
is improved over the previous work's, which ranged from 
40\% to 80\%. Most of this improvement is derived from 
the smaller statistical error on the atmospheric neutrino 
MC sample used in the present analysis.

If the $K^0_L$ search is excluded from the lifetime limit computation 
and the data set restricted to the SK-I period, 
the result is slightly worse than the previous analysis,
1.1$\times$10$^{33}$~years compared to 1.3$\times$10$^{33}$~years.
Despite the higher signal efficiencies and reduced systematic 
uncertainty on the expected background rates in this analysis, 
the increase in the number of observed candidate events in the  
$K^0_S\rightarrow\pi^+\pi^-$ Method~1 search degrades the lifetime limit. 
The three candidates found in the former analysis are also found 
here, but three additional events that were previously 
just outside of the signal region have now migrated in.
Under the older reconstruction, their momenta fell slightly 
below the cut value of 250~MeV/$c$ and thus failed to meet 
condition (B4) from Sec.~VI-A. 
The improved reconstruction has increased their estimated momenta beyond this threshold.
The $\mu^+$ candidate momentum reconstruction has been checked
using the atmospheric neutrino MC and the shift 
in the reconstructed momentum relative to the true momentum,
and the momentum resolution are both smaller in the present work. 
The new result represents an improvement to the analysis.

\section{CONCLUSION}

The SK $p\rightarrow\mu^+ K^0$ decay search presented in \cite{kk}
has been updated to include improved event reconstruction algorithms
and an additional 81.1 kton$\cdot$years of data.
A search for proton decay into $K^{0}_L$ has also 
been added. There is no significant excess of proton decay 
candidates found and the data remain consistent 
with the expected atmospheric neutrino background rates
for both the $p\rightarrow\mu^+ K^0_S$ 
and $p\rightarrow\mu^+ K^0_L$ searches.
Therefore a partial lifetime lower limit is set at 
 $\tau /B_{p\rightarrow \mu^{+} K^0 } > 1.6\times 10^{33}$~years 
at the 90\% confidence level
using 172.8 kton$\cdot$years of data.
The lifetime limits obtained by the $K^0_S$ and $K^0_L$ searches alone
are 1.37$\times$10$^{33}$ years and 0.55$\times$10$^{33}$ years,
respectively.
This result gives further constraints on 
relevant SUSY GUT models.

\section{Acknowledgments}

We gratefully acknowledge the cooperation of the
Kamioka Mining and Smelting Company.
The Super-Kamiokande experiment has been built and operated
from funding by the Japanese Ministry of Education,
Culture, Sports, Science and Technology, the United
States Department of Energy, and the U.S. National Science
Foundation. Some of us have been supported by
funds from the Korean Research Foundation (BK21),
the National Research Foundation of Korea (NRF-20110024009),
the State Committee for Scientific Research
in Poland (grant1757/B/H03/2008/35), the Japan
Society for the Promotion of Science, and the National
Natural Science Foundation of China under Grants No.10575056.



\end{document}